\documentclass[twocolumn,superscriptaddress,prl,amssym,floatfix]{revtex4}
\usepackage{graphicx}
\begin{document}

\title{Pair condensation of bosonic atoms induced by optical lattices}

\author{Mar{\'\i}a Eckholt}
\affiliation{Max-Planck-Institut f\"ur Quantenoptik,
  Hans-Kopfermann-Str.~1, Garching, D-85478, Germany.}

\author{Juan Jos\'e Garc{\'\i}a-Ripoll}
\affiliation{Facultad de CC. F{\'\i}sicas, Universidad Complutense de Madrid,
  Ciudad Universitaria s/n, Madrid, E-28040, Spain.}

\begin{abstract}
  We design a model of correlated hopping for bosonic atoms in optical
  lattices. Such model exhibits three kinds of phases, comprehending a Mott
  insulator, a charge density wave and a pair quasi-condensate. The
  implementation of the model relies on two-state atoms embedded in
  state-dependent lattices and having spin-dependent interactions. Contrary
  to other models of pairing, our design is not based on perturbative
  effects and should be observable in current experiments.
\end{abstract}
\maketitle

Since the achievement of Bose-Einstein condensation in alkali atoms
\cite{anderson95, davis95, bradley95}, we have witnessesed two major
breakthroughs in the many-body physics of cold atoms. One is the
realization of Cooper pairing and the BCS to BEC transition with
fermions \cite{regal04, zwierlein04, bourdel04}. The other one is the
implementation of lattice Hamiltonians using neutral atoms in optical
lattices \cite{greiner02,jaksch98}. Supported by this success, many
theoretical papers suggest using cold atoms with two goals: the
quantum simulation of well known Hamiltonians such as Hubbard models
\cite{jaksch98} and spin lattices \cite{duan03}, and the quest for new
physics such as bosonic quantum Hall effect \cite{wilkin00,paredes01}
and lattice gauge theories \cite{jaksch03, osterloh05}. In this work
we aim at the latter, introducing a robust mechanism of pairing that
works for bosonic atoms, opens theoretical challenges and is suited
for the recent experiments in optical superlattices
\cite{anderlini07,foelling07}.

Pairing is a central concept in strongly correlated states. In
particular, it is the essence of ordinary BCS superconductivity. In
BCS theory, an attractive interaction mediated by a phonon bath is
the basis by which electrons merge into bosonic pairs that conduct
electricity without friction. Similar phenomena appear in the BCS
pairing of fermionic atoms, where the natural attraction is enhanced
by Feschbach resonances \cite{regal04, zwierlein04, bourdel04}.

Another, less known mechanism for pairing is correlated hopping. It
appears naturally in fermionic tight-binding models \cite{foglio79,
  karnaukhov95, karnaukhov94, arrachea94, arrachea94b, deboer95,
  vidal01} and in quantum magnetism \cite{schmidt06}, consisting on the
motion of particles being influenced by the environment. This is
normally reflected by terms of the form $n_ia^{\dagger}_ja_k$
appearing in the Hamiltonian. Correlated hopping could lead to the
formation of bound electron pairs \cite{arrachea94b, arrachea94} and
it has been put forward as an explanation for high $T_c$
superconductivity \cite{hirsch89, marsiglio90}.

\begin{figure}[b]
  \centering
  \includegraphics[width=\linewidth]{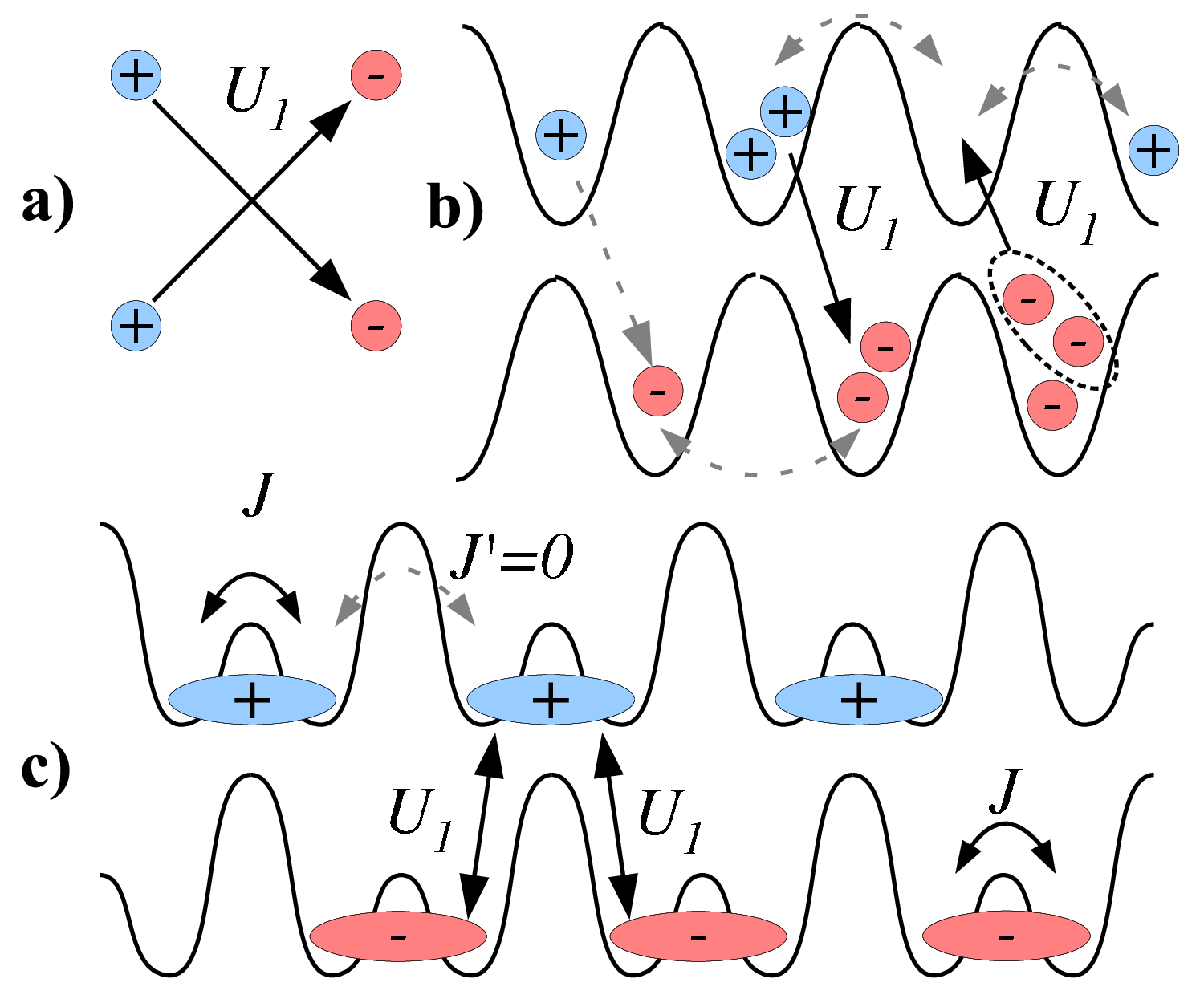}
  \caption{(a) Two atoms in state $|+\rangle$ collide and change into
    state $|-\rangle$. (b) When trapped in a state-dependent optical
    lattice, changing the state implies also jumping between lattice
    sites. We plot some forbidden (dashed) and allowed (solid)
    processes. Only pairs of atoms can hop between lattices. (c) In
    order to maximize the overlap between atoms in different states
    and thus the strength of correlated hopping, in this paper we
    consider two state-dependent superlattices.}
  \label{fig:superlattice}
\end{figure}

We will introduce a different mechanism for pairing, which is based on
transport-inducing collisions. As shown in
Fig.~\ref{fig:superlattice}, when atoms collide they can mutate their
internal state. If the atoms are placed in a state-dependent optical
lattice, whenever such a collision happens, the pair of atoms must
jump to a different site associated to their new state. For deep
enough lattices, as in the Mott insulator experiments
\cite{greiner02}, this coordinated jump of pairs of particles will be
the dominant process and the atoms will become a superfluid of pairs.

This work consists of three parts.  The first one begins with a
setup of cold atoms in a deep optical superlattice. We show how
collisions combined with state-dependent lattices lead to a strong
correlated tunnelling, as described above. The outcome is an
effective Hamiltonian which can be related to the experimental
parameters. We will then draw a realistic phase diagram using both
Gutzwiller \cite{krauth92} and Matrix Product States (MPS)
\cite{verstraete04a} variational methods. Both procedures
essentially identify three kinds of phases: a charge-density wave
(CDW), a Mott insulator (MI) and a pair condensate or pair
superfluid (PSF). In the last part of the paper we consider
different experimental procedures to detect the different insulating
and superfluid orders.

\begin{figure}
  \includegraphics[width=0.895\linewidth]{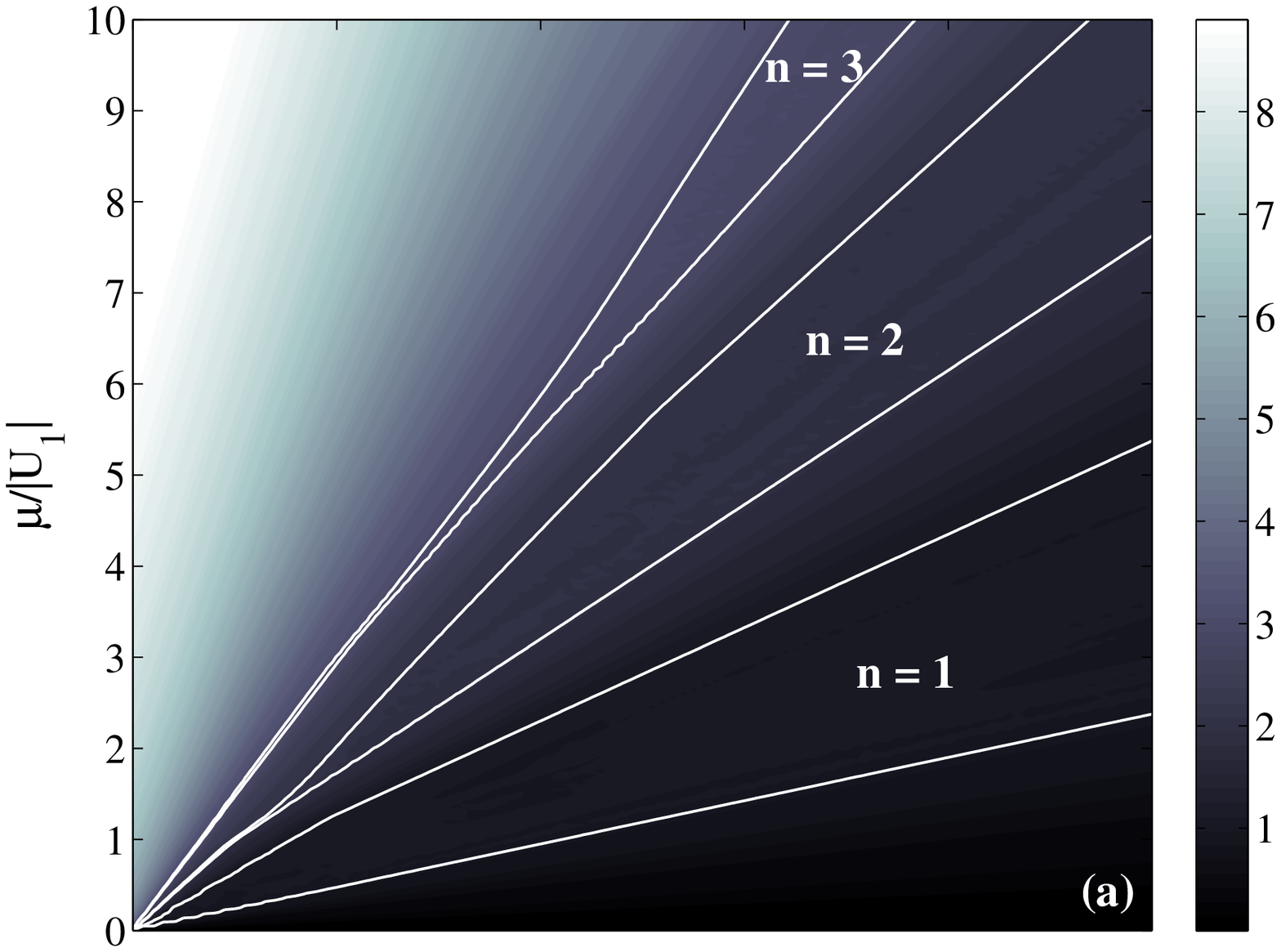}
  \includegraphics[width=0.92\linewidth]{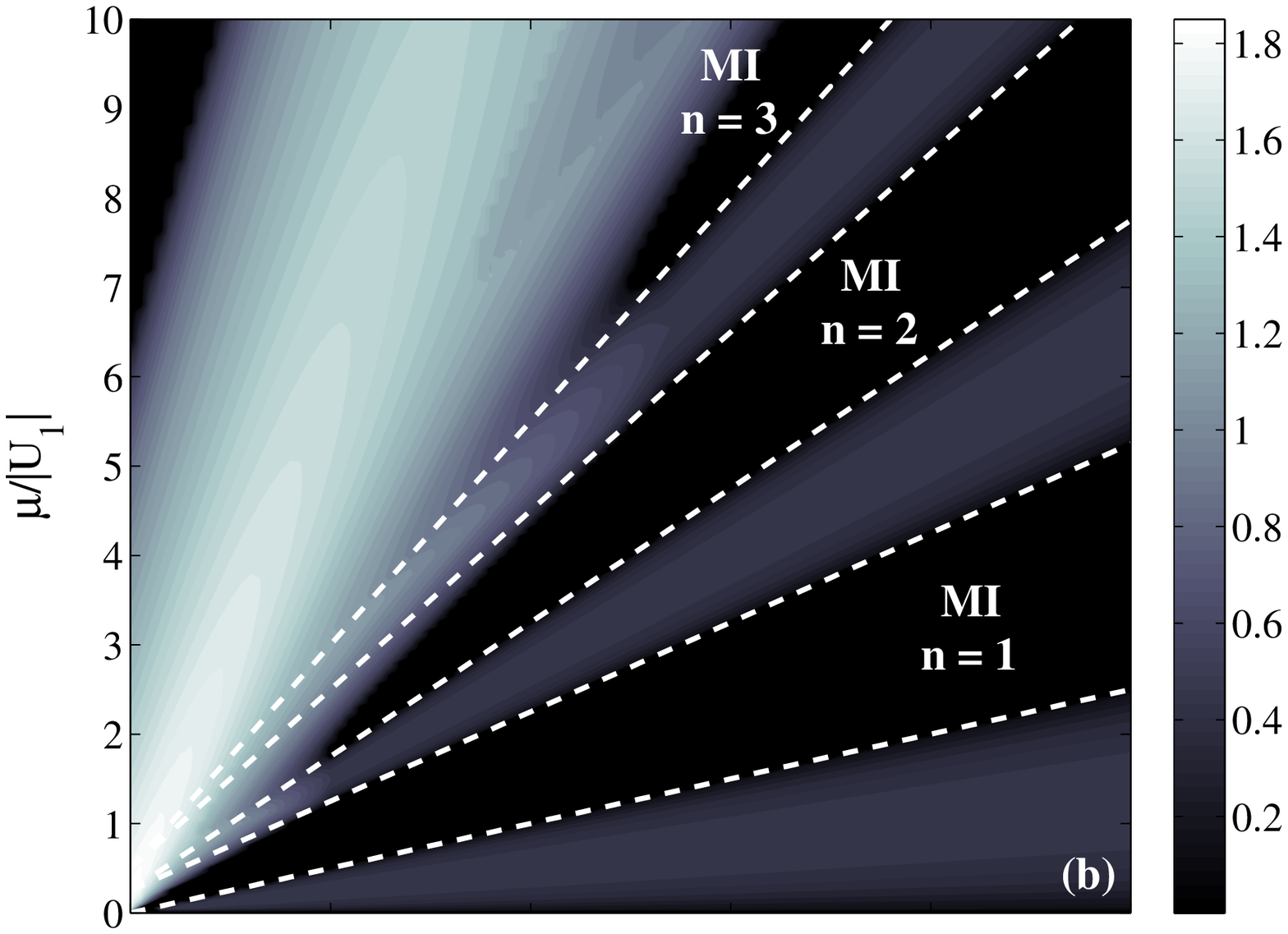}
  \includegraphics[width=0.92\linewidth]{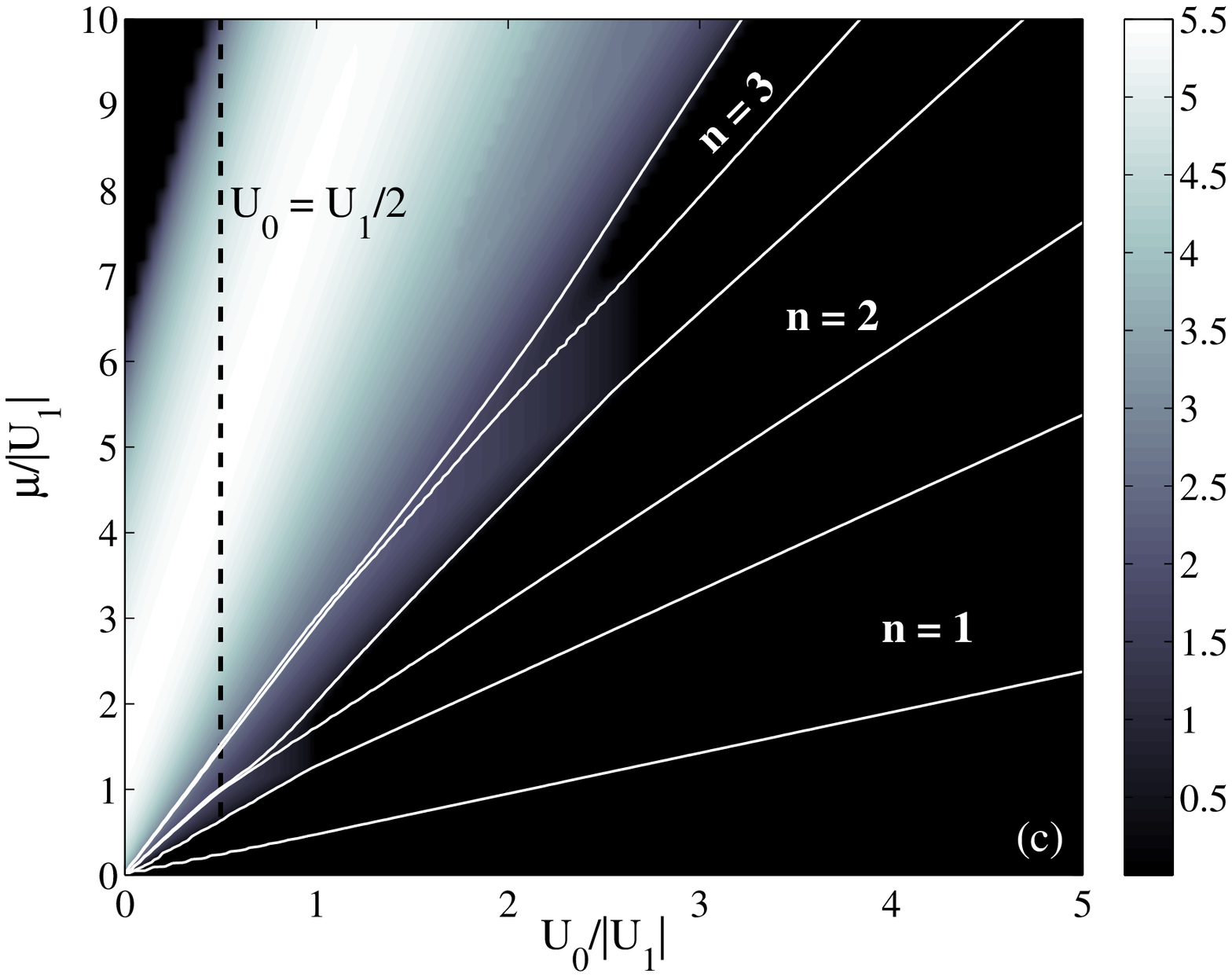}
  \caption{Ground state properties estimated with the Gutzwiller
    wavefunction (\ref{gutzwiller}). Grayscale plots of (a) the
    density, $\bar n$, (b) density fluctuations, $\Delta n$, and (c)
    pair condensate order parameter $\langle c^2\rangle$. Dashed lines
    mark the analytical estimates (\ref{mup}), while solid lines
    delimit regions of integer filling. All
    plots cover the same region $(U_0,\mu)\in [0,5U_1]\times [0,10U_1]$.}
  \label{fig:gutzwiller}
\end{figure}

Let us begin by considering the setup in
Fig.~\ref{fig:superlattice}c, where bosonic atoms in two internal
states, $a_{+}$ and $a_{-}$, are trapped in state-dependent
superlattices that spatially overlap. The hopping outside a double
well is negligible. In this limit, we have two bands of localized
states with energies proportional to the hopping amplitude on a
double well, $+J$ and $-J$. In particular the lowest band will be
given by the bosonic states $c_{2k} := \frac{1}{\sqrt{2}}(a_{2k+} +
a_{2k+1+}),$ and $c_{2k+1} := \frac{1}{\sqrt{2}}(a_{2k+1-} +
a_{2k+2-}).$

If the interaction is weak compared to the band gap $2J$, only the
coupling between states in the same band will be relevant. We will
focus on a particular situation in which the trapped states are
actually dressed states, $a_\pm := (a_\uparrow \pm
a_\downarrow)/\sqrt{2}$, and the interaction is diagonal in those
other operators, $a_\uparrow$ and $a_\downarrow$. This situation is
reached by combining the optical lattice configuration in
Ref.~\cite{garciaripoll07} with asymmetric on-site interactions
\begin{equation}
  H_k = \frac{U_0}{2} : (n_\uparrow + n_\downarrow)^2 :
  +~U_1 n_\uparrow n_\downarrow.
\end{equation}
Note that the term $U_1n_\uparrow n_\downarrow$ is not invariant under
rotations of the internal state of the atoms. When expressed in the
$|\pm\rangle$ basis, it is this term that scatters pairs of atoms from
one cloud to another [Fig.~\ref{fig:superlattice}a]. Since the
superlattices have a relative displacement, these collisions give rise
to transport. Using the approximation $U_0,U_1\ll 2J$ one arrives at
our central model
\begin{equation}
  H = \sum_{\langle i,j\rangle}\left[
    U :(n_i + n_j)^2:~+~V~n_i n_j~-~t~c_i^{\dagger 2}c_j^2
    \right]
  \label{model}
\end{equation}
where $U = (2U_0+U_1)/16$, $V = -U_1/8$, and $t = U_1/16$.

Out of these terms, $V$ favors neighboring sites to be filled, $t$ is
the correlated hopping that spreads pairs of atoms through the
lattice, and $U \geq 0$ restricts the total number of atoms preventing
collapse. If $U=0$ and $V=0$, we expect the bosons to form a
superfluid of paired particles, while for strong interactions there
should be a quantum phase transition to an insulator.

\begin{figure}
  \includegraphics[width=0.92\linewidth]{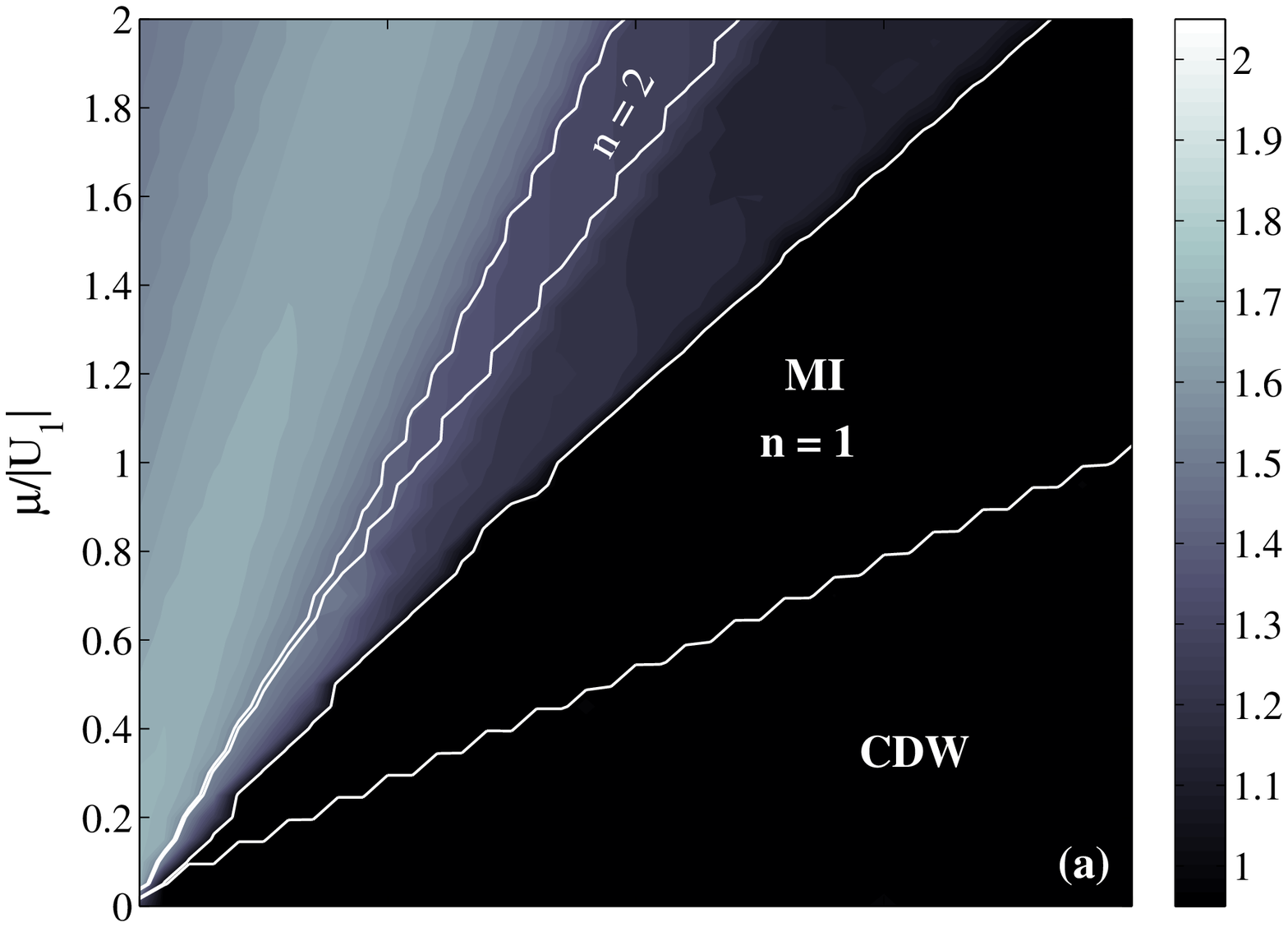}
  \includegraphics[width=0.92\linewidth]{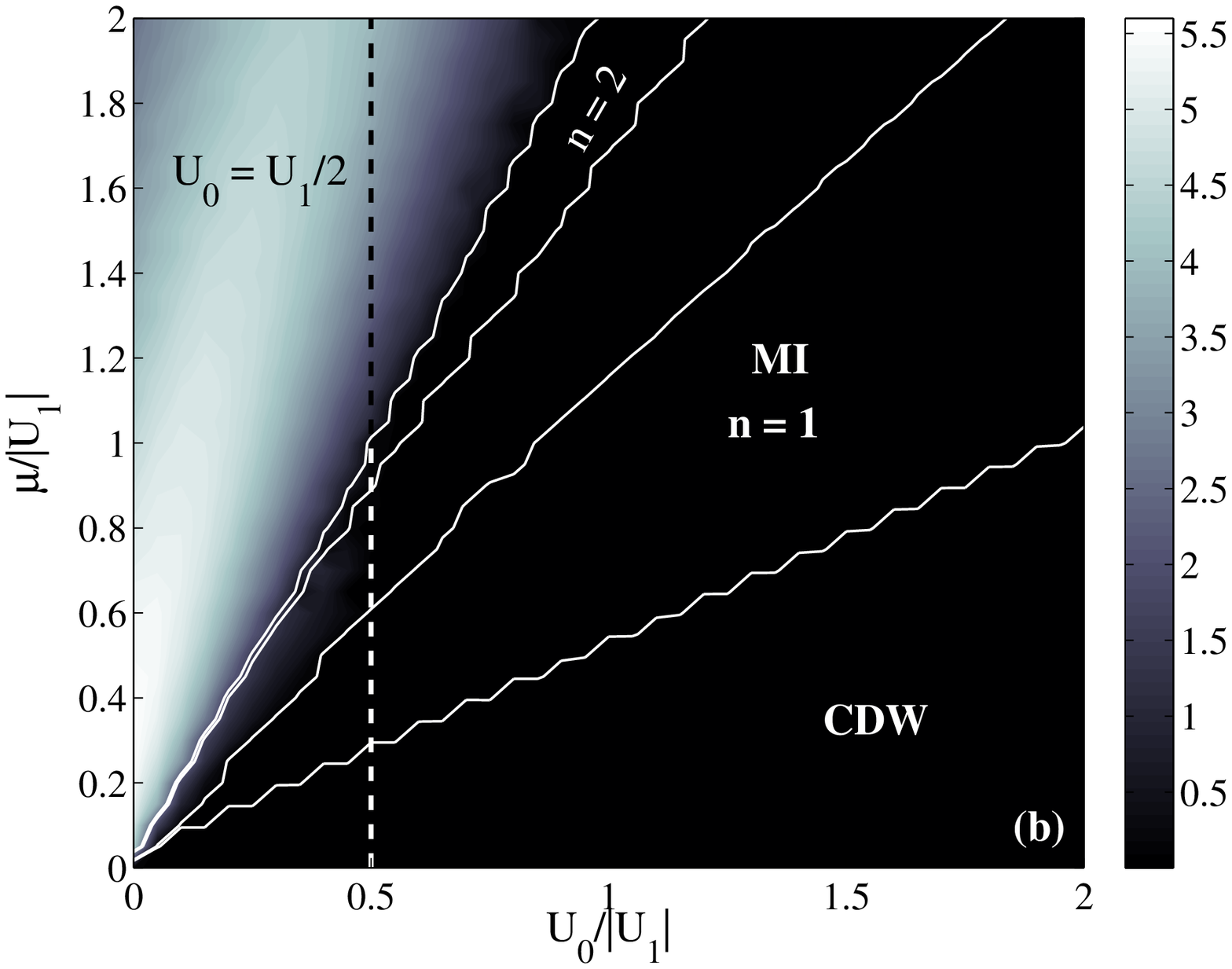}
  \caption{Ground state properties estimated with the MPS method. We
    plot (a) the fluctuations of the density, $\Delta n$, and (b) the
    averaged two-particle correlator, $\sum_\Delta |C^2_\Delta| / L$,
    where $C^2_\Delta :=\langle c^{\dagger
      2}_ic^2_{i+\Delta}\rangle$. Solid lines delimit regions of
    integer filling. All
    plots cover the same region $(U_0,\mu)\in [0,2U_1]\times [0,2U_1]$.}
  \label{fig:mps}
\end{figure}

The first evidence of pairing is obtained by studying a lattice with
only two particles. This setup has two potential ground states: one
with both particles isolated and zero energy, and a delocalized
state of a single pair $|\psi_{\mathrm{pair}}\rangle \propto \sum_i
c_i^{\dagger 2} |0\rangle$ with energy $E_2=2(U-V)$. Therefore, if
$U< V$ or $U_0<U_1/2$, the pair of bosons will be bound, $E_{2} <
0$. An important question is whether this criterion survives in the
presence of other atoms, or if many-body effects may destroy the
pairing.

In order to study the many-body physics of our model we have used the
Gutzwiller ansatz, which is a variational method based on the product
state \cite{krauth92}
\begin{equation}
  |\psi_{\mathrm{GW}}\rangle =
  \prod_i \sum_n \frac{1}{\sqrt{n!}}f_n c_i^{\dagger n} |0\rangle
  \label{gutzwiller}
\end{equation}
Minimizing the expectation value of the free energy, $F := H - \mu N$,
with respect to the variables $f_n$ under the constrain of fixed norm,
$\sum |f_n|^2 = 1$, one obtains the phase diagram in the phase space
of interaction and chemical potential, $(U_1,U_0,\mu)$. For the sake of
simplicity we have used $U_1:=1$ as unit of energy everywhere.

The results are shown in Fig.~\ref{fig:gutzwiller}, where we plot the
expectation values of the density, $\bar n = \langle n \rangle =
\langle c^\dagger c\rangle$, the particle number variance, $\Delta n
:= \sqrt{\langle n^2 - \bar n^2\rangle}$, and what we identify as the
order parameter of the paired superfluid, $\langle c^2\rangle$.  From
the zeros of the density fluctuations we recognize the Mott regions
with $\bar n = 1, 2$ and $3$ particles per site. The boundary of the
region with $\bar n = 1$ can be estimated analytically using a
Gutzwiller state with nonzero components $f_0$, $f_1$ and $f_2$,
which gives
\begin{equation}
  \mu_p = \frac{U_0}{2},\; \mathrm{and}\;
  \mu_h = U_0 + \frac{1}{4}.
  \label{mup}
\end{equation}
These are the energies to add a particle or to make a hole in this
insulating lobe. If $\mu < \mu_h$, the Gutzwiller ansatz gives
fractional densities but, as we will see below, this is an artifact of
the uniform trial wavefunction. If $\mu > \mu_p$ and $U_0 < U_1/2$, we
obtain a region of nonzero $\langle c^2\rangle$. We take this as a
sign of a long-range coherence in the two-body density matrix which
cannot be attributed to an ordinary Bose-Einstein condensate, since
the same simulation gives $\langle c\rangle = 0$

The Gutzwiller ansatz does not reproduce accurately neither the
location of phase transitions nor the behavior of correlators. For
instance, $C^1_\Delta :=\langle c_i^\dagger c_{i+\Delta}\rangle$ and
$C^2_\Delta :=\langle c_i^{\dagger 2} c^2_{i+\Delta}\rangle$ computed
with Eq.~(\ref{gutzwiller}) are both uniform functions. In order to
study these properties, we have used the MPS method
\cite{verstraete04a} to estimate variationally the ground state of our
model. The MPS is a more complex wavefunction that, for large enough
computational resources and not too large systems, should reproduce
the correlations in the superfluid regime. Calculations with up to
$40$ sites reveal that matrices of size $D=40$ are enough to pinpoint
the different phases.

In Fig.~\ref{fig:mps} we plot the same quantities as for the
Gutzwiller ansatz. The contour plots reveal that the $\bar n=1$ and
$\bar n=2$ insulating regions still exist, with a similar size and
shape as in the Gutzwiller case. As we explain later, the region above
$\mu_p$ again shows a paired superfluid behavior, which extends even
beyond the limit $U_0 = U_1/2$ where the binding energy of an isolated
pair becomes positive [Fig.~\ref{fig:mps}b]. The biggest difference is
in the triangle $0 \leq \mu < \mu_h$. In this region of low density,
the bosons arrange forming a charge density wave, i.~e. a pattern
alternating 0 and 1 atom per site.

\begin{figure}[b]
  \includegraphics[width=0.85\linewidth]{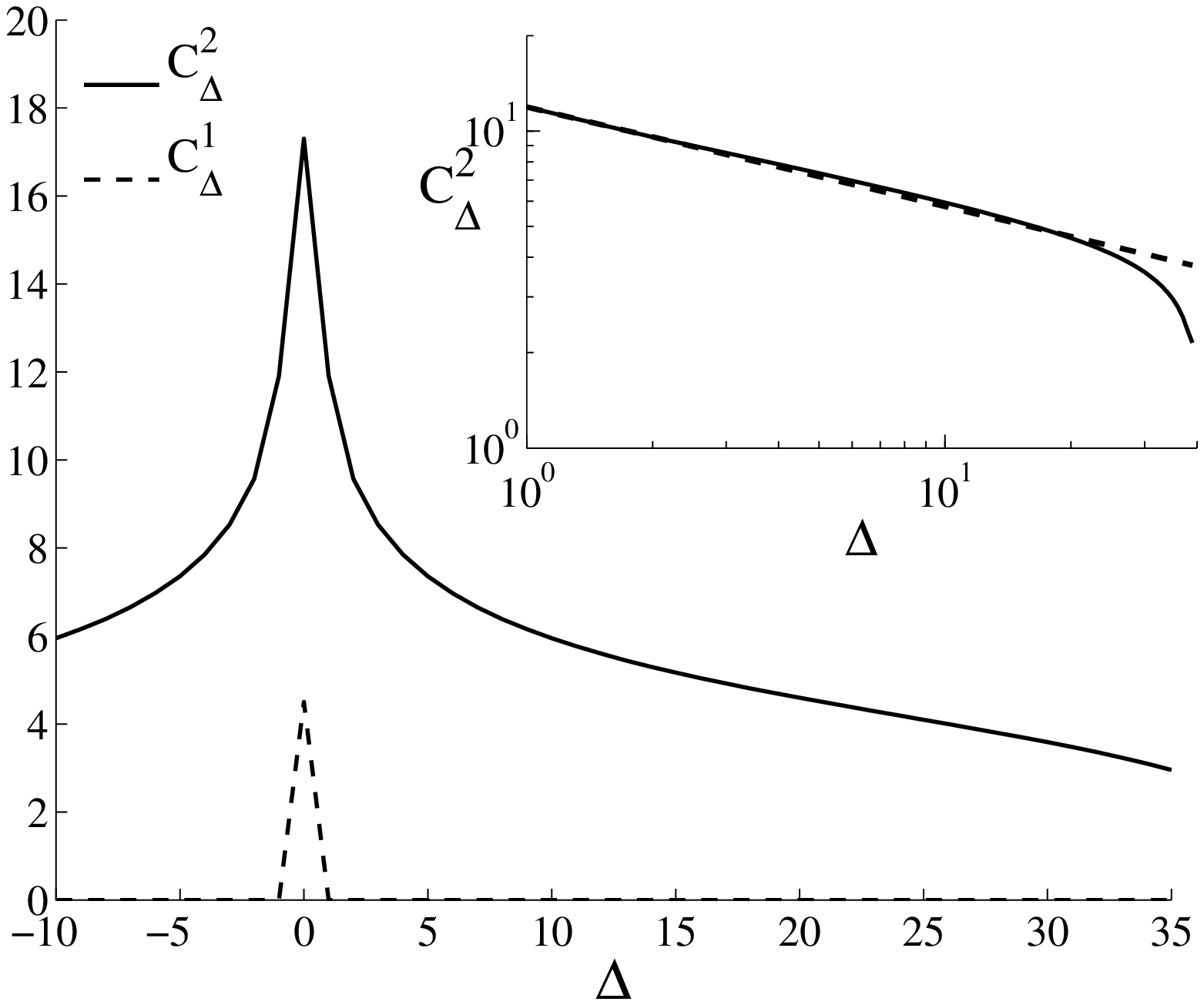}
  \caption{Single particle, $C^1_\Delta:=\langle
    c^\dagger_ic_{i+\Delta}\rangle$ (dashed), and two particle,
    $C^2_\Delta:=\langle a^{\dagger2}_ia^2_{i+\Delta}\rangle$,
    correlator (solid) for $\mu/U_1=0.9, U_0/U_1=0$ and a lattice
    with 40 sites. The inset repeats the plot for $C^2_\Delta$ in
    log-log scale, with a fit $C^2_\Delta \propto \Delta^{0.32}$ (dash).}
  \label{fig:corr}
\end{figure}

Regarding the superfluid, quasi-long-range order is identified by a
slow decay of the off-diagonal elements of the single- and
two-particle density matrix. As shown in Fig.~\ref{fig:corr}, the
single particle correlator $C^1_\Delta=\bar \delta_{\Delta,0}$ is only
different from zero at $\Delta=0$, where it becomes the density. This
could have indicated the presence of a Mott phase, were it not for the
nonzero value of $\Delta n$ (Fig.~\ref{fig:mps}b) and of the two
particle correlator, $C^2_\Delta$, that decays slowly at long
distances. Due to the size of our simulations, we have not yet been
able to determine the behavior of $C^2_\Delta$, but numerical fits of
curves like Fig.~\ref{fig:corr} suggest an algebraic decay,
$C^2_\Delta \sim \Delta^{-\alpha}$, with a nonuniversal exponent
around $\alpha < 0.7$ that depends on $U_0$ and $\mu$.

Let us now introduce another possible experimental setup, illustrated
in Fig.~\ref{fig:superlattice}b and where two lattices trapping the
two different species are shifted by half a period or $\lambda /4$. As
usual, the trapped states and the interaction are described in
different basis. One can expand the interacting Hamiltonian in terms
of Wannier functions together with operators defining the trapped
states. This leads to Eq.~(\ref{model}) with: $U = (2U_0+U_1)/4$, $V =
-(2\beta U_0+ U_1)/2$ and $t = -U_1/4$.  Here, $U_0$ and $U_1$ are the
on-site interactions in the atomic basis and the factor $\beta = \int
|w(x)|^2|w(x-\lambda/4)|^2dx / \int |w(x)|^4dx\ll 1$, computed using
the Wannier functions $w(x)$, measures the relative strength of
interaction between different lattices with respect to those on the
same lattice site.

The phases that appear in our setup can all be recognized
experimentally. First of all, the MI and the CDW have both a well
defined number of particles per site and no coherence. Their time of
flight pictures will have no interference fringes \cite{greiner02}
and the noise correlation will exhibit peaks at certain momenta
\cite{altman04,foelling05}. In addition, the CDW corresponds to a
setup where either $n_{+}$ or $n_{-}$ are uniformly zero. Finally,
the energy gap in these insulators can be probed by static
\cite{greiner02} or spectroscopic means \cite{stoeferle04}.

Regarding the pair superfluid, it is a perfect ``conductor'' with a
gapless excitation spectrum. Lacking single-particle order,
$C^1_\Delta \sim 0~a.e.$, it will also not show interference patterns
in the time of flight images. In order to measure $C^2_\Delta$ and
detect the pairing, we suggest to use photoassociation and to build
molecules out of pairs of atoms. Since the molecules will be built
on-site, the nonzero correlator $C^2_\Delta$ will translate into
long-range order for the molecules, which should now exhibit an
interference pattern in time-of-flight images, slightly blurred due to
the phase fluctuations inherent to 1D.

Finally, all energy scales are referenced to the asymmetry of the
interactions, $U_1$. Therefore, the pair binding energy can be
potentially as strong as the strength of the on-site interaction in a
typical Mott insulator \cite{greiner02}, about 1 kHz. In
practice, while the most commonly used atoms such as Rubidium have
smaller asymmetries, they can be enhanced using Feschbach resonances.

Summing up, we have introduced a mechanism by which bosonic atoms with
repulsive interactions can exhibit correlated hopping and pairing. The
model (\ref{model}) exhibits multiple phases, among which the most
relevant is a superfluid of paired bosons. All phases are connected by
second order quantum phase transitions and can be produced and
identified using variations of current experiments
\cite{anderlini07,foelling07}. These ideas can be generalized to
higher dimensions and other other kinds of interaction.

We thank Miguel Angel Mart{\'\i}n-Delgado for useful
discussions. M.E. acknowledges support from the CONQUEST project.
J.J.G.R acknowledges financial support from the Ramon y Cajal Program
of the Spanish M.E.C., from U.S. NSF Grant No. PHY05-51164 and from
the spanish projects FIS2006-04885 and CAM-UCM/910758.

\end{document}